\documentclass[11pt]{article}

\usepackage[authoryear]{natbib}

\usepackage{graphicx}
\usepackage{caption}
\usepackage{subcaption}
\usepackage{todonotes}
\usepackage{cancel}
\usepackage{float}
\usepackage{multirow}

\usepackage{color}
\usepackage{soul}
\usepackage[utf8]{inputenc}
\usepackage[T1]{fontenc}  

\usepackage{algorithm} 
\usepackage{algpseudocode}
\usepackage{authblk}
\usepackage{mathtools}
\usepackage[margin=1.1in]{geometry}
\usepackage{amssymb}
\usepackage{amsmath}
\usepackage{amsthm}
\usepackage{siunitx}
\usepackage[capitalise]{cleveref}
\usepackage{amsthm}
\usepackage{calc}

\usepackage{times}

\newcommand{\D}{\mathbf{ D}}
\renewcommand{\S}{\mathbf{ S}}

\newcommand{\n}{\mathbf{ n}}

\newcommand{\grad}{\nabla}

\newcommand{\uvec}{\mathbf{ u}}
\newcommand{\vvec}{\mathbf{ v}}

\newcommand{\fgref}[1]{Fig.~\ref{#1}}

\newcommand{\yr}{yr}

\renewcommand{\eqref}[1]{~(\ref{#1})}

\begin{document}

\title{Increasing stable time-step sizes of the free-surface problem arising in ice-sheet simulations}

\author[1]{Andr\'{e} L\"{o}fgren}
\author[1,2]{Josefin Ahlkrona \thanks{Corresponding author\\ Email adresses: ahlkrona@math.su.se (J.Ahlkrona), andre.lofgren@math.su.se (A. L\"{o}fgren) christian.helanow@math.su.se (C. Helanow)}}
\author[1,2]{Christian Helanow}

\affil[1]{Department of Mathematics, Stockholm University, Stockholm, Sweden}
\affil[2]{Swedish e-Science Research Centre (SeRC), Stockholm, Sweden}

%
%
%

\maketitle

\section*{Abstract}
Numerical models for predicting future ice-mass loss of the Antarctic and Greenland ice sheet requires accurately representing their dynamics. Unfortunately, ice-sheet models suffer from a very strict time-step size constraint, which for higher-order models constitutes a severe bottleneck since in each time step a nonlinear and computationally demanding system of equations has to be solved. 

In this study, stable time-step sizes are increased for a full-Stokes model by implementing a so-called free-surface stabilization algorithm (FSSA). Previously this stabilization has been used successfully in mantle-convection simulations where a similar, but linear, viscous-flow problem is solved. 

By numerical investigation it is demonstrated that instabilities on the very thin domains required for ice-sheet modeling behave differently than on the equal-aspect-ratio domains the stabilization has previously been used on. Despite this, and despite the nonlinearity of the problem, it is shown that it is possible to adapt FSSA to work on idealized ice-sheet domains and increase stable time-step sizes by at least one order of magnitude. The FSSA presented is deemed accurate, efficient and straightforward to implement into existing ice-sheet solvers.


\section{Introduction}\label{sec:intro}

Our warming climate affects the dynamics of the ice sheets on Antarctica and Greenland, causing an increased ice discharge into the oceans. The meltwater contributes significantly to global sea-level rise \citep{IPCC5,IPCCspecialreport}. Accurate computer models of ice dynamics are a crucial tool in order to confidently predict future ice-mass loss; unfortunately, such models suffer from instabilities that severely restrict feasible simulation lengths.

Modeling the evolution of an ice sheet amounts to solving a very viscous, non-Newtonian, gravity-driven moving-boundary problem. The standard approach for solving this moving-boundary problem is to: 1) compute the ice velocity and pressure by solving the nonlinear full-Stokes equations or some approximation thereof, and 2) use the velocity to update the surface position by solving the so-called free-surface equation. This process is then repeated in each time step, and effectively couples the free-surface equation to the full-Stokes equations. A limitation of this approach is that it suffers from instabilities unless the time-step size is small. However, the alternative - a fully implicit time-stepping scheme - involves solving for velocity and pressure several times in each time step in an iterative scheme, which is computationally very expensive.  
\begin{figure}[h!]
 \centering
      \includegraphics[width=0.9\textwidth,trim={1cm 0 0 1cm},clip]{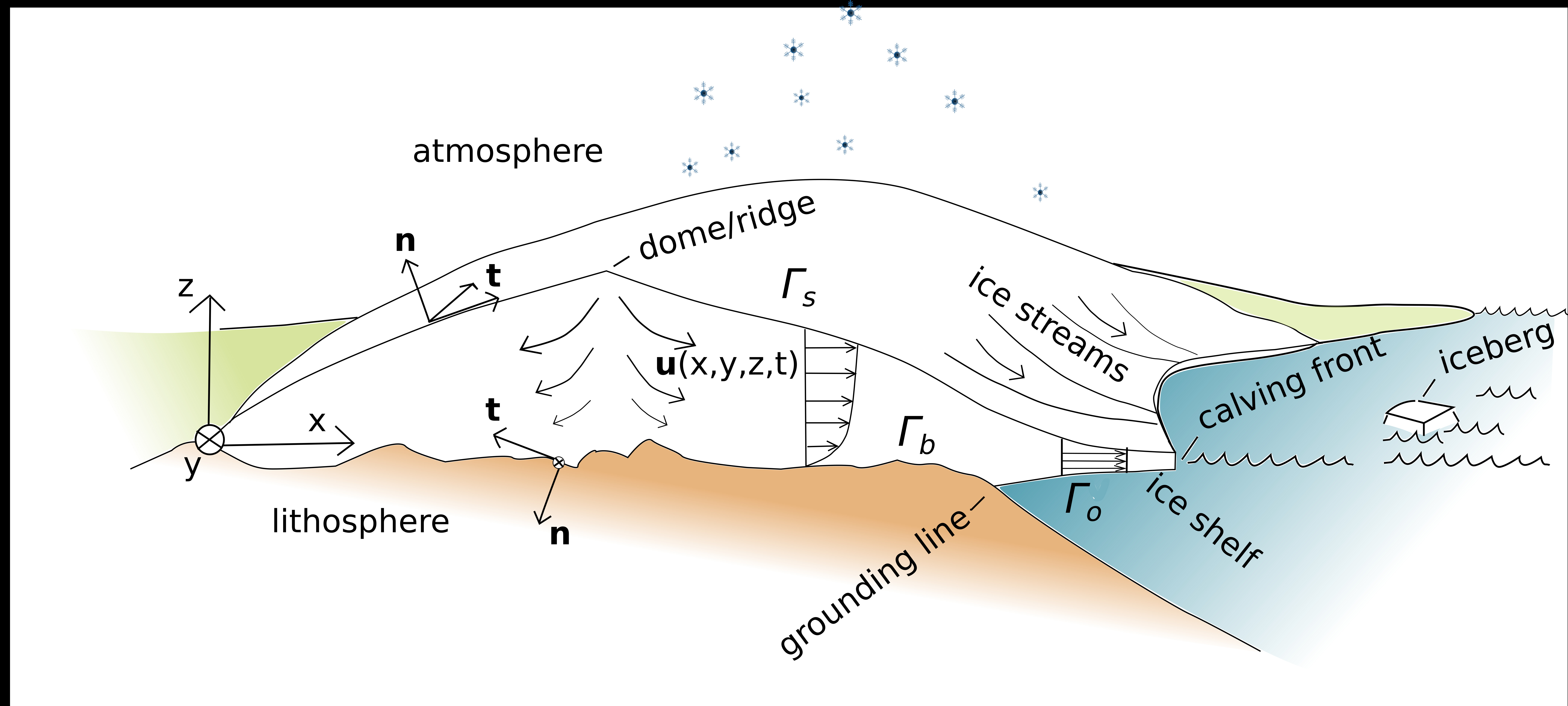}
		\caption{A cross section of an ice sheet (white) on an underlying bedrock (brown) and an adjacent ocean (blue). The position of the ice-sheet surface $\Gamma_s$ depends on the velocity of the ice $\mathbf{u}$.} \label{fig:icesheet}
\end{figure}

To our knowledge there are no successful attempts at ameliorating this type of instability for full-Stokes ice-sheet models. Stabilization is sometimes discussed for the type of instabilities arising in advection-dominated problems  (e.g., \citep{gmd-13-6425-2020, gmd-2020-397}) and saddle-point problems (e.g., \citep{HelanowAhlkrona}) but here the argument is made that the most severe instabilities are rather related to the way the coupling between velocity and surface position is handled and that these instabilities can be mitigated.

The connection between the velocity-surface coupling and stability properties is poorly understood for full-Stokes models. Studies on time-step size restrictions in ice-sheet models have only been made for a very crude approximation to the full-Stokes equations called the shallow-ice approximation \citep{Hindmarsh2001,hindmarsh_payne_1996,BuelerBrown2009,GREVE2002649}. For the shallow-ice approximation it is possible to explicitly state how the velocity depends on the position of the free surface, rewrite the free-surface equation and derive a parabolic time-step size constraint $\Delta t < C \Delta x^2$ using Fourier analysis. Here $\Delta t$ denotes the time-step size and $\Delta x$ the mesh resolution. 

For the full-Stokes equations, velocities cannot be determined explicitly as a function of the free surface and for this reason it is not possible to derive a time-step size constraint in the same way as can be done for the shallow-ice approximation. However, numerical experiments show that the full-Stokes-coupled free-surface equation also suffers a strict constraint on the time-step size \citep{CHENG201729}. 

In this paper the nature of the instabilities in full-Stokes ice-sheet simulations are investigated and a stabilization which increases the largest stable time-step size (LST) by at least one order of magnitude without compromising accuracy is introduced. This is achieved by transferring and modifying methods used in mantle-convection modeling, where the issue of the restrictive time-step size has successfully been overcome by introducing a pseudo-time-stepping scheme known as free-surface stabilization algorithm (FSSA) \citep{ANDRESMARTINEZ201541,ROSE201790,KAUS2010}. 

Mantle convection is also modeled as highly viscous free-surface flow \citep{ANDRESMARTINEZ201541,ROSE201790,KAUS2010}. However, ice-sheet dynamics involves a few complications that are not taken into account in previous implementations of FSSA. Firstly, ice viscosity varies with strain rate and becomes very high at locations where strain rates are low, whereas the Earth's mantle in the works of \citep{ANDRESMARTINEZ201541,ROSE201790,KAUS2010} is modeled as a Newtonian fluid. Secondly, ice-sheet domains are highly anisotropic in the sense that they are much thinner than they are wide, while the modeling domains in, e.g., \citep{KAUS2010} are isotropic. Thirdly, the ice surface is subject to precipitation and melting, leading either to an accumulation or ablation of mass on the surface. Lastly, ice-sheet domains are also dome shaped with steeply inclined surfaces at their fronts, while domains in mantle convection problems are effectively extending infinitely in the lateral direction with low surface gradients.

The remainder of the paper is structured as follows: Sect.~\ref{sec:equations} presents the equations governing the flow of ice. The semi-implicit time discretization and its inherent stability issues are then discussed in Sect.~\ref{sec:timediscretization}. The FSSA method is introduced and discussed in Sect.~\ref{sec:stabilization}. Sect.~\ref{sec:experiments} explores 
stability properties of ice-flow by numerically investigating how the LST depends on geometry, choice of Euler method and whether or not FSSA is used. To this end, simulations are run on three different set-ups: The first is an adaptation of a benchmark experiment \citep{ANDRESMARTINEZ201541} to the regime of ice-sheet modeling. For this type of domain FSSA have been shown to increase the LST, for a linear mantle rheology. The second experiment considers a prognostic two-dimensional ice-sheet benchmark called EISMINT \citep{Huybrechts}, modeling the build-up of an ice sheet subject to snow accumulation. The third experiment is a three-dimensional prognostic experiment of a Vialov profile \citep{vialov1958}. Lastly, in Sect.~\ref{sec:summary} the outcome of the experiments and the main findings are summarized, along with a discussion of the outlook of FSSA in the context of ice-sheet modeling.


\section{Governing Equations}\label{sec:equations}

\subsection{Governing Equations}

The velocity $\uvec=(u_x,u_y,u_z)$ and the pressure $p$ of an ice sheet are given as the solution to the full-Stokes equations

\begin{alignat}{2}\label{eq:pStokes}
\left.\begin{matrix}
&-\S (\D\uvec) + \grad p &= \mathbf{f}  \\ 
& \grad \cdot  \uvec &=0
\end{matrix}\right\} \quad \text{ in } \Omega.
\end{alignat}
Here the domain $\Omega \subset \mathbb{R}^d \mathbb{d}, d=2,3$, is bounded by the ice-atmosphere interface (the surface) $\Gamma_s$, ice-ocean interface $\Gamma_o$ and ice-bedrock interface $\Gamma_b$, so that the boundary of the domain $\partial \Omega = \Gamma_s \cup \Gamma_o \cup \Gamma_b$; see \fgref{fig:icesheet} for an illustration of the ice-sheet domain. The body force $\mathbf{f} = -\rho g \hat{\mathbf{z}}$ is due to gravity and drives the ice flow.
$\S(\D\uvec)$ is the deviatoric stress tensor and depends on the strain-rate tensor $\D\uvec:=\frac{1}{2}(\grad \uvec + \grad(\uvec)^T)$ as
\begin{equation}
\S(\D\uvec)=2\mu(\D\uvec) \D\uvec.
\end{equation}
Here $\mu$ is the non-constant effective viscosity of ice which through the constitutive equation of ice, commonly called Glen's flow law, is related to the strain rate as
\begin{equation}
\mu(\D\uvec)=\frac{1}{2}A^{-1/n}(\epsilon_{crit}^2+\frac{1}{2}||\D\uvec||^2_F)^{\frac{1/n-1}{2}}.
\label{eq:glen}
\end{equation}

The dependency of the viscosity on the strain rate is due to the non-Newtonian material properties of ice and is what sets the equations apart from the linear Stokes equations. For ice, the stress exponent $n$ in Glen's flow law, Eq.~\eqref{eq:glen}, is set to $n=3$, which makes the viscosity very large for small strain rates. To prevent the viscosity from approaching infinity at zero strain rates, a small regularization constant called the critical shear rate $\epsilon_{crit}$ is added to the numerical implementation of Glen's flow law. The viscosity may also depend on temperature via the rate factor $A$, but here isothermal conditions are considered such that $A$ is constant. 

The non-Newtonian properties of ice flow makes the full-Stokes equations computationally expensive to solve. This is mainly for two reasons: firstly because a Newton or Picard iteration has to be employed in order to linearize the equations, meaning the full-Stokes equations~\eqref{eq:pStokes} (see Alg.~\ref{alg:solproc}) has to be solved repeatedly in each time step. Secondly, it causes the condition number of the linearized discrete system to be high, which slows down the convergence of the linear solvers.

For well-posedness of the full-Stokes equations, boundary conditions have to be specified on all parts of the boundary. In this study, land-terminating ice sheets are considered so that $\Gamma_o = \emptyset$. On the surface $\Gamma_s$, atmospheric pressures and wind stresses are assumed to be negligible compared to internal stresses in the ice sheet. Under this assumption the surface satisfies the stress-free boundary condition~\eqref{bc:stress-free}. At the remainder of the boundary, $\partial \Omega \setminus \Gamma_s$, either a no-slip condition~\eqref{bc:no-slip} is imposed, or an impenetrability condition in the normal direction~\eqref{bc:impenetrability} together with a free-slip condition in the tangential direction~\eqref{bc:free-slip}.
\begin{align}
(-p \mathbf{I} + 2\mu \D)\cdot \mathbf{n} &= \mathbf{0}, \quad &\text{on } \Gamma_s \label{bc:stress-free} \\
\uvec&=\mathbf{0}, &\text{on } \partial \Omega \setminus \Gamma_s \label{bc:no-slip}\\
\nonumber &\textnormal{ or }&\\
\quad \uvec \cdot \mathbf{n} &= 0, &\text{on } \partial \Omega \setminus \Gamma_s \label{bc:impenetrability} \\
\mathbf{t}_i \cdot \mathbf{\S(\D\uvec)} \cdot \mathbf{n} &= 0. &\text{on } \partial \Omega \setminus \Gamma_s\label{bc:free-slip}
\end{align}
Here $\mathbf{I}$ denotes the identity matrix, $\mathbf{t}_i$ are tangent vectors spanning the plane orthogonal to the outward pointing unit normal $\mathbf{n}$. 

The time evolution of the surface boundary is described by a separate equation called the free-surface equation
\begin{equation}
\frac{\partial h}{\partial t}+\uvec_H \cdot \grad_H h = u_z + a_s , \text{on } \Gamma_s,
\label{eq:free_surface}
\end{equation}
where $h(x,y,t)$ is the height of the surface $\Gamma_s$ above sea level at a certain horizontal coordinate $(x,y)$ at time $t$. The horizontal part of the velocity $\uvec$ is given by $\uvec_H=(u_x,u_y)$ in three dimensions and $\uvec_H=u_x$ in two dimensions, where $\grad_H h = \left ( \frac{\partial h}{\partial x}, \frac{\partial h}{\partial y} \right)$ and $\grad_H h = \frac{\partial h}{\partial x}$ denotes surface gradients in the respective dimension. The surface mass balance $a_s$ is the rate at which snow accumulates (or ablates) perpendicular to the surface. 


\section{Time Discretization and Instability}\label{sec:timediscretization}

In this section a time-stepping scheme that is standard in glaciology is reviewed, along with a discussion of its (in)-stability properties. It uses the following type time-discretization of the free-surface equation~\eqref{eq:free_surface}

\begin{equation}
		\frac{h^{k+1} - h^k}{\Delta t } + \mathbf{u}_H^k \cdot \grad_H h^{k+\gamma} =  u_z^k +a_{s}^{k+\gamma},
	\label{eq:discreteicesurface}
\end{equation}
where $\gamma=0$ results in a forward Euler discretization, and $\gamma=1$ a backward Euler discretization. The standard time-stepping approach is to in each time step $t^k$ solve the full-Stokes equation on the domain $\Omega^k$ for $\uvec^k$, and then solve the discretized free-surface equation~\eqref{eq:discreteicesurface} with $\uvec^k$ as coefficients to obtain the new domain $\Omega^{k+1}$, see Alg.~\ref{alg:solproc}.
\begin{algorithm}[h!]
\caption{General Solution Procedure}
\label{alg:solproc}
\begin{algorithmic}[1]
\For {each time step $t^k$}
\State Solve the full-Stokes equation for ice velocity $\mathbf{u}^k$ and pressure $p^k$
\State Insert $\mathbf{u}^k$ into the discretized free surface equation~\eqref{eq:discreteicesurface}
\State Solve Eq.~\eqref{eq:discreteicesurface} for the new surface $h^{k+1}$
\State Update the computational mesh using $h^{k+1}$
\State Let $k=k+1$
\EndFor
\end{algorithmic}
\end{algorithm}

This method is relatively efficient as it avoids solving the computationally expensive nonlinear full-Stokes equations more than once in each time step. However, the total number of times the full-Stokes equations are solved for in a prognostic simulation is still proportional to the time-step size and constitutes a bottleneck as the method is prone to numerical instabilities that severely restrict stable time-step sizes. The aforementioned computational cost of solving the nonlinear full-Stokes equations makes it infeasible to perform long-term simulations for time-step sizes conforming to the stability constraint. However, stable time-step sizes are usually much smaller than the natural time scale at which an ice sheet evolves and increasing time-step sizes through stabilization can thus be done without compromising the accuracy of the solver. 

Implicit methods are generally more stable than explicit methods. Therefore a first attempt at alleviating stability issues might be to use backward Euler time discretization. However, backward Euler for $h$ alone does not constitute a fully-implicit method since that would also require using the velocities from the next time step, $\uvec^{k+1}$. In numerical experiments simulating mantle convection it has been demonstrated that using only velocities from the current $\mathbf{u}^{k}$ give rise to a so-called sloshing instability for which backward Euler is to no avail \cite{KAUS2010}. 

The core of the matter lies in the coupling between the free-surface equation and the full-Stokes equations; the solution $(\uvec,p)$ of the full-Stokes equations depends on the shape of the domain $\Omega$ (i.e., on the surface position $h$), and the velocity $\uvec$ enters as a coefficient in the free-surface equation.  

\begin{figure}[H]
			\centering
				\includegraphics[width=0.7\linewidth]{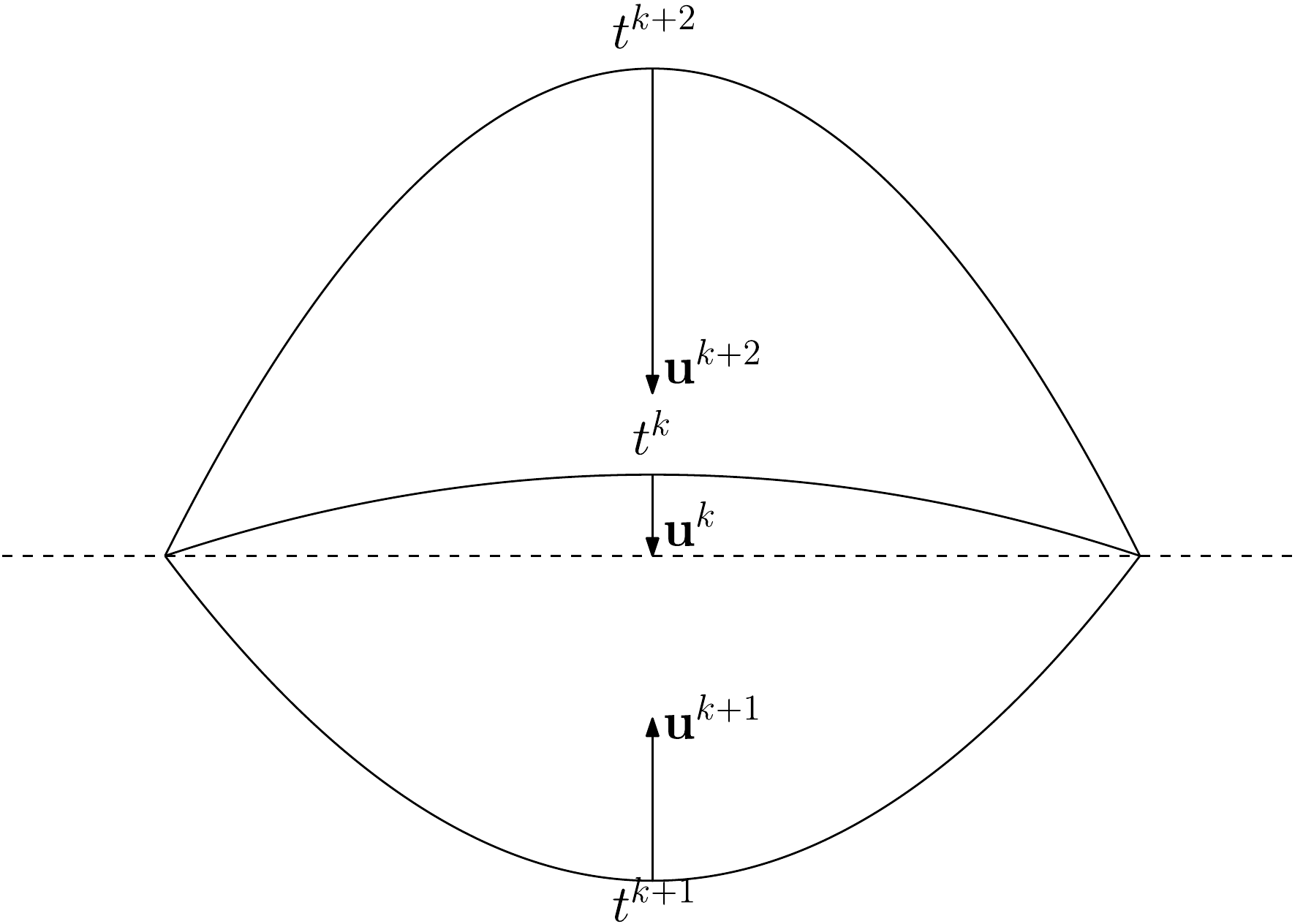}
				\caption{A free surface shown at times $t^k$, $t^{k+1}$ and $t^{k+2}$ when updated using velocities $\uvec^k$ of the current time step $t^k$. This causes the surface to overshoot its equilibrium configuration, the dashed line, such that it ends up in the configuration shown at time $t^{k+1}$. At this stage it obtains even larger velocities $\uvec^{k+1}$, as indicated by the length of the arrow, which are then used to update the surface. This causes the surface to overshoot even more in time step $t^{k+2}$, resulting in instability.}
			\label{fig:FSSA_explained}
		\end{figure}

The following contrived example serves to demonstrate how using $\mathbf{u}^k$ instead of $\mathbf{u}^{k+1}$, even for backward Euler discretization, gives rise to sloshing and how a fully-implicit solver (with respect to $\uvec$) in theory ameliorates it. Consider a free surface at time $t^k$ positioned as shown in Fig.~\ref{fig:FSSA_explained}. The gravity driven nature of the flow leads the surface to flatten out over time and approach its equilibrium configuration, the dashed line in Fig.~\ref{fig:FSSA_explained}. This results in lower and lower velocities, so that $||\uvec^{k+1}|| < ||\uvec^{k}||$. However, if $\uvec^{k}$ is used as a coefficient in the free-surface equation~\eqref{eq:discreteicesurface} this can results in a too large change $h^{k+1} - h^k$ such that the surface overshoots its equilibrium configuration and ends up in the position shown at $t^{k+1}$ in Fig.~\ref{fig:FSSA_explained}. In this case the surface has obtained a larger velocity than in the previous time step, causing the surface to overshoot the equilibrium even more at time step $t^{k+2}$, leading to instability. If instead at time $t^k$ the velocity $\uvec^{k+1}$ is solved for and used for updating the surface, it will never overshoot equilibrium since the magnitude of the surface velocities decreases over time, such that if a large time-step size is used the magnitude of velocities $\uvec^{k+1}$ will also be small.

The sloshing instability has been studied theoretically by \cite{ROSE201790}. An explicit expression for the LST is derived under the assumption that the free surface is a small perturbation around a flat, hydrostatic surface. The surface can then be viewed as a superposition of exponentially decaying normal modes, with each mode having a decay time $\tau_i$ and its time evolution described by the linear test equation



\begin{equation}
		\frac{\partial h_i(x,t)}{\partial t} = \frac{\partial \tilde{h_i}(x)e^{-t/\tau_i}}{\partial t}  =-\frac{h_i(x,t)}{\tau_i} \approx -\uvec_H \cdot \grad_H h + u_z.
\label{eq:}
\end{equation}

To determine the decay time \citet{ROSE201790} solve an eigenvalue problem. However, here a simpler approach is taken and the argument is made that if the initial surface is a sinusoid it consists of a single normal mode, such that the minimum decay time can be approximated as $\tau_{min} \approx z_0/||-\uvec_H \cdot \grad_H h + u_z||_{\infty}$, where $z_0$ is the height of the surface above its equilibrium position and $\tau_{min}$ is the minimum decay time of all normal modes. The $LST=2\tau_{min}$ since a stable surface is allowed to travel at maximum from $z_0$, past the equilibrium surface, to the other side up until $-z_0$. This amounts to a distance of $2z_0$. The LST is thus

\begin{equation}
		 LST \approx \frac{2 z_0}{||-\uvec_H \cdot \grad_H h + u_z ||_{\infty}} \approx \frac{2 z_0}{||u_z ||_{\infty}}.
	\label{eq:tau_min}
\end{equation}

A central question is whether sloshing, or a similar type of instability, also occurs when these methods are used in an ice-dynamical setting, and to which extent it affects the stability constraint on the time-step size. This is one of the key issues addressed in this paper and for this purpose it is investigated under what circumstances the LST estimate Eq.~\eqref{eq:tau_min} holds.


\section{The FSSA Method}\label{sec:stabilization}

As mentioned in the previous section, the sloshing instability could in theory be remedied by using a fully implicit method with respect to $\uvec$ when solving the free-surface equation. However, solving implicitly for $\uvec^{k+1}$ requires solving the computationally expensive nonlinear full-Stokes equations multiple times in each time step, which is infeasible for long-term simulations where many time steps have to be performed. 

Following the works of \citet{KAUS2010}, instabilities are instead mitigated by altering the Stokes equations to compute an estimate $(\tilde{\uvec}^{k+1},\tilde{p}^{k+1})$ of $(\uvec^{k+1},p^{k+1})$. The stabilization is developed within a framework using the finite element method (FEM), and is based on finding a solution to the weak formulation of the full-Stokes equations~\eqref{eq:pStokes}: Find $(\uvec^k,p^k) \in \boldsymbol{\mathcal{X}} \times \mathcal{Q}$ such that

\begin{alignat}{2}\label{eq:weak-cont1}
\int_{\Omega^k}\S(\D\uvec^k) : \grad \vvec - \int_{\Omega^k} p^k  \grad \cdot \vvec &= \int_{\Omega^k}\mathbf{f} \cdot \vvec \quad &\forall \vvec \in \boldsymbol{\mathcal{X}}, \\
\int_{\Omega^k}\grad \cdot \uvec^k q &=0 \quad &\forall q \in \mathcal{Q}.
\label{eq:weak-cont2}
\end{alignat}
The natural spaces for the continuous weak solution \citep{belenki2012} are the Sobolev and Lebesgue spaces
\begin{align*}
  \boldsymbol{\mathcal{X}} & := \mathbf{W}^{1,\frac{n+1}{n}}_0(\Omega^k) := \{ \uvec^k \in [W(\Omega^k)^{1,\frac{n+1}{n}}]^d :  \uvec^k = \mathbf{0} \text{ or } \uvec^k \cdot \n = 0 \text{ on }  \partial \Omega^k \setminus \Gamma_s^k\},\\
  \mathcal{Q} & := L^{n+1}(\Omega^k).
\end{align*}

To mimic solving for $\mathbf{u}^{k+1}$, the right hand side of Eq.~\eqref{eq:weak-cont1} is modified to be an approximation of $(\mathbf{f}, \mathbf{v})_{\Omega^{k+1}}$. This makes the full-Stokes equations "aware" of the spatially evolving domain by estimating the impact of the force of gravity at the next time step.	Evaluating $(\mathbf{f}, \mathbf{v})_{\Omega^{k+1}}$ at time step $k$ is obviously not possible since it would require knowing the domain $\Omega^{k+1}$ in advance; however, by applying Reynolds transport theorem to $(\mathbf{f}, \mathbf{v})_{\Omega^{k+1}}$ it can be expressed purely in terms of integrals over the domain $\Omega^k$ 

\begin{equation}
\int_{\Omega^{k+1}}\mathbf{f}\cdot \mathbf{v} \approx \int_{\Omega^k}\mathbf{f} \cdot \mathbf{v} + \theta \Delta t \int_{\partial \Omega^k}(\mathbf{u} \cdot \mathbf{n}) (\mathbf{f} \cdot \mathbf{v}).
\label{eq:Taylor}
\end{equation}
Here a stabilization parameter $\theta \in \mathcal{R}^{+}$ has been included in Reynolds transport theorem. The role of $\theta$ is that it controls the "implicitness" of the solver, where $\theta=0$ gives an explicit solver (with respect to $\uvec$) and $\theta=1$ is quasi implicit.

The body force integral in Eq.~\eqref{eq:weak-cont1} is now replaced by the expression above, yielding the original stabilized weak formulation of \cite{KAUS2010}: Find $(\tilde{\uvec}^{k+1},\tilde{p}^{k+1}) \in \boldsymbol{\mathcal{X}} \times \mathcal{Q}$ such that:

\begin{alignat}{2}\label{eq:discrete_stab1}
  \nonumber \int_{\Omega^k} \S(\D \tilde{\uvec}^{k+1}) : \grad \vvec &- \int_{\Omega^k} \tilde{p}^{k+1} \grad \cdot \vvec - \theta \Delta t \int_{\partial \Omega^k} ( \tilde{\uvec}^{k+1} \cdot \mathbf{n} ) (\mathbf{f}\cdot \mathbf{v} )\quad &\\
  &= \int_{\Omega^k}\mathbf{f}\cdot \vvec \quad &\forall \vvec \in \boldsymbol{\mathcal{X}}, \\
\int_{\Omega^k}\grad \cdot \tilde{\uvec}^{k+1} q &=0 \quad &\forall q \in \mathcal{Q}.
\end{alignat}
Empirically, this weak formulation has been shown, for a linear mantle rheology, to increase stable time-step sizes significantly without compromising accuracy nor efficiency \citep{KAUS2010}. The benefit of using this method compared to a fully implicit time-stepping scheme is that the full-Stokes equations only have to be solved once in each time step.

Typically, ice sheets evolve in part by the addition or removal of mass to the surface.
The effect of the accumulation rate $a_s(x, y, t)$ on the surface movement can be taken into account in the FSSA if the surface velocities $\mathbf{u}$ used in the Reynolds transport theorem~\eqref{eq:Taylor} are replaced by $\mathbf{u} \rightarrow \mathbf{u} + a_s \hat{\mathbf{z}}$. 

The result of this substitution is to add an additional term $\theta \Delta t \int_{\Gamma_s^k}  (a_s \hat{\mathbf{z}} \cdot \mathbf{n} ) (\mathbf{f}\cdot \mathbf{v})$ to the right hand side of Eq.~\eqref{eq:discrete_stab1}, which finally gives the stabilized weak formulation for ice flow: Find velocity and pressure $(\tilde{\uvec}^{k+1},\tilde{p}^{k+1}) \in \boldsymbol{\mathcal{X}} \times \mathcal{Q}$ such that

\begin{alignat}{2}\label{eq:discrete_stab2}
\nonumber \int_{\Omega^k} \S(\D \tilde{\uvec}^{k+1}) : \grad \vvec &- \int_{\Omega^k} \tilde{p}^{k+1} \grad \cdot \vvec - \theta \Delta t \int_{\Gamma_s^k} ( \tilde{\uvec}^{k+1} \cdot \mathbf{n} ) (\mathbf{f}\cdot \mathbf{v} )\quad &
\\ &= \int_{\Omega^k}\mathbf{f}\cdot \vvec  +  \theta \Delta t \int_{\Gamma_s^k}  (a_s \hat{\mathbf{z}} \cdot \mathbf{n} ) (\mathbf{f}\cdot \mathbf{v}) \quad &\forall \vvec \in \boldsymbol{\mathcal{X}}, \\
\int_{\Omega^k}\grad \cdot \tilde{\uvec}^{k+1} q &= 0 \quad &\forall q \in \mathcal{Q}.
\end{alignat}
In the above expression, the boundary integral $\partial \Omega^k$ has been replaced by an integral over the ice surface $\Gamma^k_s$, by employing the impenetrability condition~\eqref{bc:impenetrability} and the no-slip boundary condition~\eqref{bc:free-slip}. 
%


\section{Numerical Experiments}\label{sec:experiments}

This section presents three different numerical experiments. Each experiment is designed to give a better understanding of the numerical (in-)stability properties of full-Stokes ice-sheet simulations, and to assess how well the free-surface stabilization algorithm (FSSA) works under ice-sheet-like conditions.

Experiment 1 is an adaptation of the two-dimensional (2D) sloshing-sinusoid experiments used in the mantle-convection studies {in \cite{ANDRESMARTINEZ201541, ROSE201790}. The purpose of Experiment 1 is to illustrate how instabilities behave on a geometrically anisotropic (relevant for ice-sheet simulations) domain compared to on a isotropic domain (as used in mantle-convection simulations in \cite{ANDRESMARTINEZ201541, ROSE201790}), and to determine whether FSSA can ameliorate instabilities in both cases. 

Experiment 2 is the classical 2D EISMINT benchmark developed by \citet{Huybrechts}, commonly used in numerical ice-sheet modeling. The purpose of this experiment is to test the stabilization in a realistic setting relevant to ice-sheet simulations.
Lastly, Experiment 3 the stabilization is tested on a three-dimensional (3D) domain. The performance of the stabilization method in 3D is most relevant to realistic ice-sheet settings, as it is for these computationally expensive simulations that better stability properties would be most beneficial.

To spatially discretize the full-Stokes equations using FEM, the domain $\Omega$ is divided into non-overlapping triangles (or tetrahedrons in 3D) to form a mesh $\mathcal{T}_h$. In Experiment 1 and 2, the \emph{inf-sup} stable Taylor-Hood element (quadratic polynomials for the velocity and linear polynomials for the pressure) are used for stabilizing the resulting saddle-point problem. In Experiment 3, due to the high computational cost of using the higher-order Taylor-Hood element in 3D, equal-order linear elements are used for both velocity and pressure. Such elements are not \emph{inf-sup} stable and need to be stabilized by other means. In this case the Galerkin least-squares stabilization is used: this type of stabilization is commonly used in ice-sheet simulations \cite{ElmerDescrip,VarGlaS,HelanowAhlkrona}. See \cite{HelanowAhlkrona} for a discussion on the numerical properties of various stabilization methods in the context of ice-sheet modeling.

The free-surface equation is discretized and solved on the boundary of the mesh used for the full-Stokes equations.  
The mesh is then updated by linearly interpolating between the new nodal positions on the free surface and the nodes on the bedrock. For ease of implementation the solution space is restricted to piecewise linear functions. 

In all experiments isothermal conditions are assumed and use the physical parameters shown in Table~\ref{tab:ice_parameters}. 

\begin{table}[H]
        \centering
        	\caption{Physical parameters and relative nonlinear-iteration tolerance $\epsilon$ given in units of (km, \yr{}, MPa), where \yr{} is year and MPa is megapascal.}
        \begin{tabular}{l | l}
		Quantity & Value in units of km, \yr{} and MPa \\
        \hline \hline
        $\rho$ & $9.137 \times 10^{-13}~$ \yr$^2$ km$^{-2}$ MPa \\
        $g$ & $9.760 \times 10^{12}~$km \yr$^{-2}$ \\
        $A$ & 100 MPa$^{-3}$ \yr$^{-1}$ \\
        $n$ & $3$ \\
		$\epsilon$ & $10^{-6}$ 
       \end{tabular}
       \label{tab:ice_parameters}
\end{table}


\subsection{Experiment 1.1: Illustration of instabilities and the effect of FSSA}

\subsubsection{Setup}
In this experiment a body of ice with a sinusoidal free surface, subject to no accumulation (i.e., $a_s=0$), residing on top of a flat bedrock is considered. As time progresses the surface relaxes until equilibrium is reached and it is completely flat. The analytical expression for the initial free surface of the domain is given by
\begin{equation}
h(x) = L_z + z_0 \cos{\left ( \frac{\pi x}{L_x} \right )},
\end{equation}
where $z_0=0.1$ km is the amplitude of the sinusoid, $L_x$ the horizontal extent and $L_z$ the height. The boundary conditions enforced are no-slip~\eqref{bc:no-slip} on the bedrock, the stress-free condition~\eqref{bc:stress-free} on the surface and free-slip~\eqref{bc:free-slip} along with an impenetrability condition in the normal direction on the sides~\eqref{bc:impenetrability}. 

\begin{figure}[H]
\center
\includegraphics[width=0.5\linewidth]{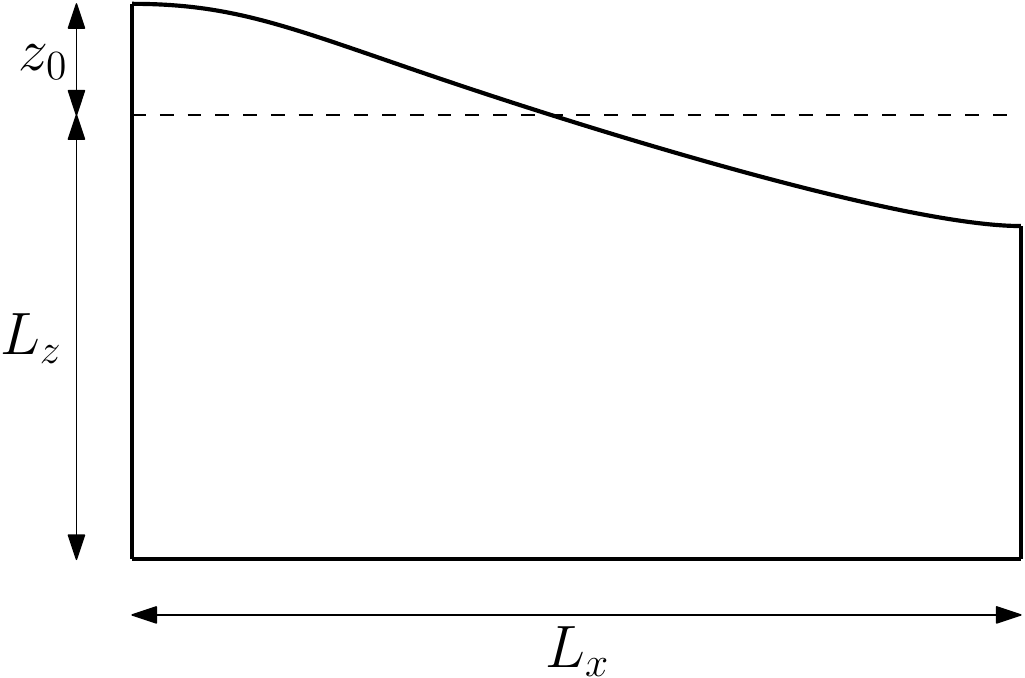}
\caption{Setup of the decaying sinusoid experiment. A free surface in the shape of a half sinusoid, overlying a flat bedrock, is allowed to relax. The sinusoid has an amplitude $z_0$ and the box has dimensions $L_x$ and $L_z$, where $L_z$ is the height of the box at equilibrium.}
\label{fig:sinusoid}
\end{figure}

This experiment is performed on two different types of domains in order to investigate the effect of geometrical anisotropy on the instabilities and to determine whether FSSA can be used to alleviate stability issues in both cases. The first setup considers an isotropic domain (see Fig.~\ref{fig:isotropic_sloshing}) with $L_x = L_z = 1$ km and the other an anisotropic domain (see Fig.~\ref{fig:anisotropic_sloshing}) with $L_x = 100$ km and $L_z = 1$ km. The isotropic domain is discretized with a horizontal resolution $n_x=30$ and a vertical resolution $n_z=30$, and the anisotropic domain with resolutions $n_x=100$ and $n_z=10$.

Four simulations (a-d) are run for each domain. Simulations (a-c) use no stabilization ($\theta = 0$) while simulation (d) is stabilized with FSSA ($\theta=1$). Simulation (a) is aimed at determining how the free surface behaves for the unstabilized formulation when a small enough time-step size is used to render a stable surface ($\Delta t = 0.01$ \yr{} in Fig.~\ref{fig:isotropic_sloshing}a and $\Delta t = 35$ \yr{} in Fig.~\ref{fig:anisotropic_sloshing}a). Simulation (b) shows how the surface behaves when it is close to being unstable ($\Delta t = 0.1$ \yr{} in Fig.~\ref{fig:isotropic_sloshing}b and $\Delta t = 70$ \yr{} in Fig.~\ref{fig:anisotropic_sloshing}b). Simulation (c) shows how the surface behaves when the time-step size is just large enough to render instabilities ($\Delta t = 0.11$ \yr{} in Fig.~\ref{fig:isotropic_sloshing}c and $\Delta t = 75$ \yr{} in Fig.~\ref{fig:anisotropic_sloshing}c). Lastly, Simulation (d) shows how the surface behaves when FSSA is used for the same or a larger time-step size causing instabilities in Simulation (c) ($\Delta t = 0.11$ \yr{} in Fig.~\ref{fig:isotropic_sloshing}d and $\Delta t = 100$ \yr{} in Fig.~\ref{fig:anisotropic_sloshing}d). In all simulations forward Euler is used (i.e., $\gamma=0$ in Eq.~\eqref{eq:discreteicesurface}).
Figures~\ref{fig:isotropic_sloshing}a and~\ref{fig:anisotropic_sloshing}a show that using a stable time-step size the surfaces approach equilibrium as expected on both domains.

\subsubsection{Results \& Discussion}

\begin{figure}[H]
\centering
\includegraphics[width=\linewidth]{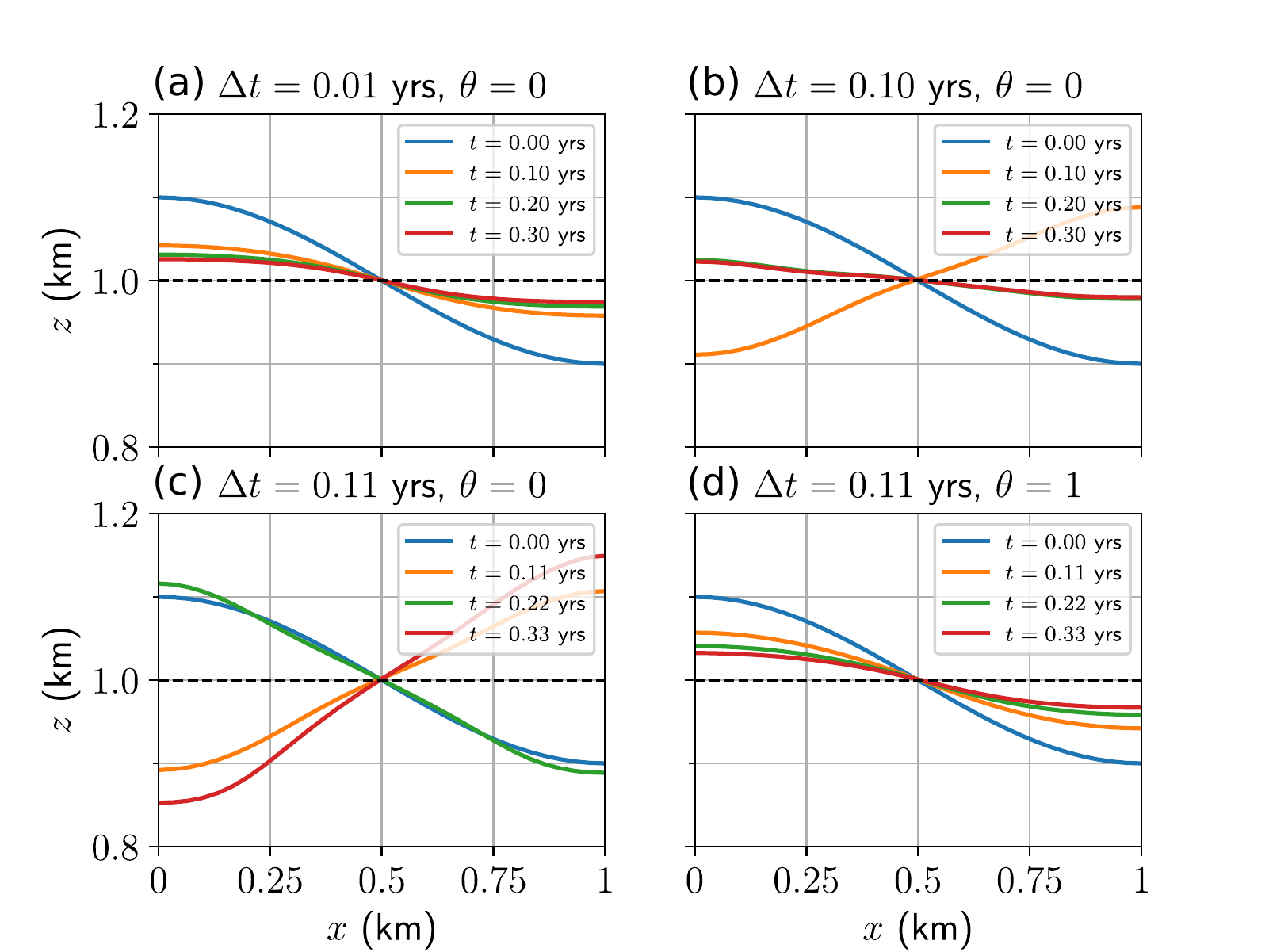}
		\caption{Surface evolution on a geometrically \textbf{isotropic} domain using different time-step sizes without FSSA (a-c) and with FSSA (d). (a) Stable time-step size $\Delta t = 0.01$ \yr{} $ \ll LST$. (b) Close to unstable time-step size $\Delta t = 0.10 $ \yr{} $< LST$. (c) Unstable time-step size $\Delta t = 0.11$ \yr{} $ > LST$. (d) Stable time-step size $\Delta t = 0.11$ \yr{} $ > LST$. In the unstable case (c), the surface is sloshing around the equilibrium position (black dashed line), with an increasing amplitude. The same sloshing behavior is seen in (b), but in this case the solver is stable since the surface approaches equilibrium. (d) Adding FSSA mitigates the sloshing instability. For this case, Eq.~\eqref{eq:tau_min} yields a theoretical LST of $0.125$ \yr{}.}
\label{fig:isotropic_sloshing}
\end{figure}

\begin{figure}[H]
\centering
\includegraphics[width=\linewidth]{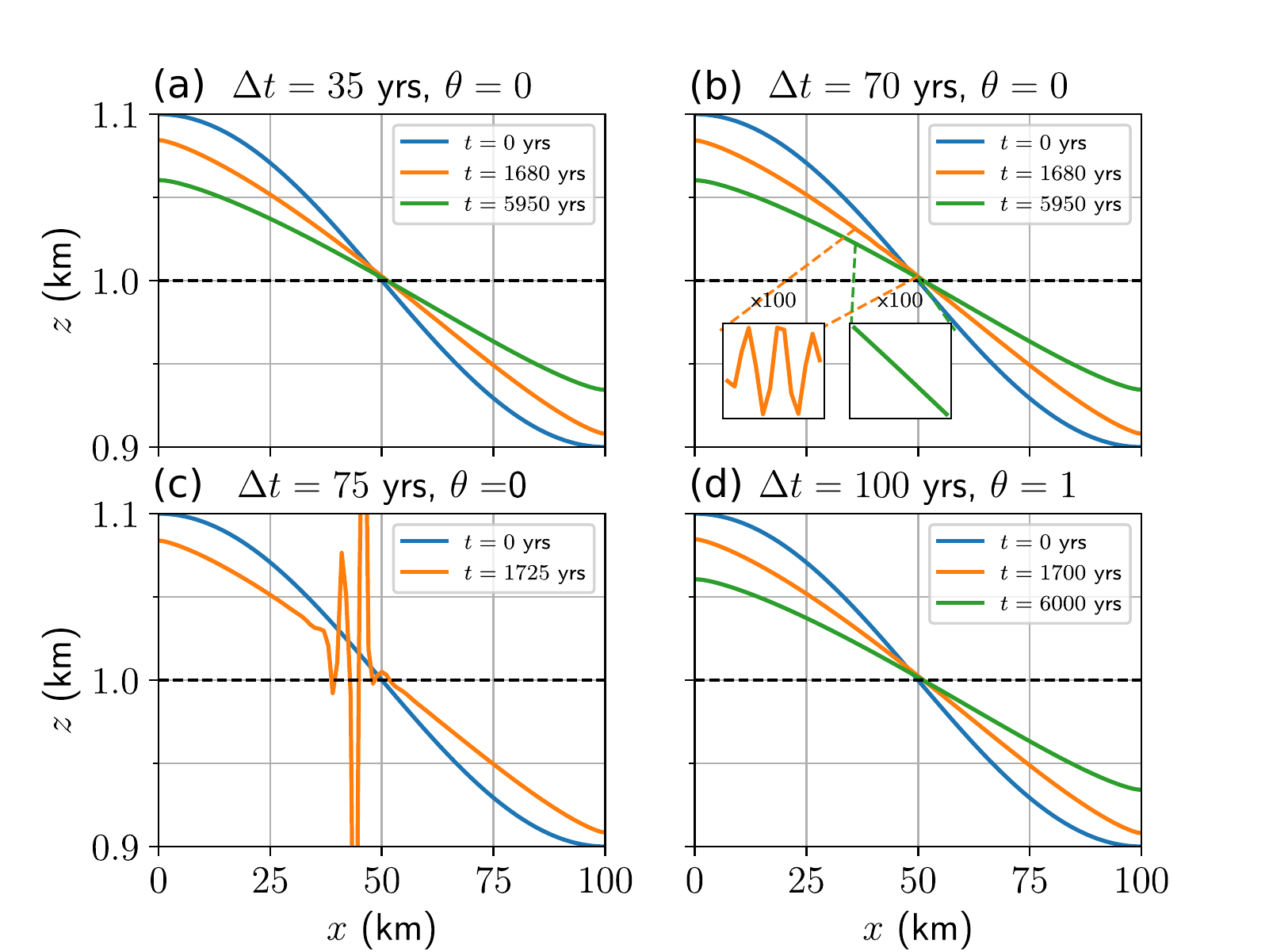}
		\caption{Surface evolution on a geometrically \textbf{anisotropic} domain using different time-step sizes, without FSSA (a-c) and with FSSA (d). (a) Stable time-step size $\Delta t = 35$ \yr{} $\ll LST$. (b) Close to unstable time-step size $\Delta t = 70$ \yr{} < $LST$. (c) Unstable time-step size $\Delta t = 75$ \yr{} $ > LST$. (d) Stable time-step size $\Delta t = 100$ \yr{}. In (b) small numerical oscillations are observed at $t=1680$ \yr{}, seen in the left sub figure. The amplitude of the oscillations have been amplified by a factor of 100 in order to make them visible. The green line shows that the oscillations eventually disappear and hence this case is stable. In (c), which uses a slightly larger time-step size, the oscillations do not disappear and instead the method diverges. (d) Adding FSSA mitigates the instability. For this case, Eq.~\eqref{eq:tau_min} yields a theoretical LST of $5200$ \yr{}.}
\label{fig:anisotropic_sloshing}
\end{figure}

The largest time-step size which renders a stable solution on the isotropic domain is, $LST=0.109$ \yr{}. This agrees well with theory, since $||-\uvec_H \cdot \grad_H h + u_z ||_{\infty} \approx 1.6$, giving a theoretical LST of $2 \cdot 0.1/1.6=0.125$ according to Eq.~$\eqref{eq:tau_min}$. On the isotropic domain sloshing is seen to be present for a time-step size slightly smaller than the $LST=0.106$ \yr{} , i.e., the surface overshoots to the other side, but it still approaches equilibrium, see Fig.~\ref{fig:isotropic_sloshing}b. The solution is thus stable but inaccurate. If the time-step size is instead slightly larger than the LST (see Fig.~\ref{fig:isotropic_sloshing}c) the surface is unstable in the sense that overshooting becomes progressively worse and equilibrium is never reached. This is the same behavior that has been observed in previous experiments for a linear mantle rheology \citep{KAUS2010,ANDRESMARTINEZ201541,ROSE201790}. Thus, on an isotropic domain, the instabilities for the nonlinear full-Stokes equations used in glaciology behave similarly to those for the linear Stokes equations.

For the anisotropic case, it can be noticed that the numerically determined $LST=72$ \yr{} is much larger than for the isotropic domain, and that the instabilities are of a different character. On the anisotropic domain the theoretical LST is, given that $||-\uvec_H \cdot \grad_H h + u_z ||_{\infty} \approx 0.000038$ m/\yr{}, as large as $5200$ \yr{}. The instabilities, which appear as early as at 72 \yr{}, do not appear to be of sloshing type.
A time-step size slightly smaller than the LST gives rise to small oscillations in the surface, as can be seen from the orange line in Fig.~\ref{fig:anisotropic_sloshing}b, which shows the free surface after 1680 \yr{}. However, in this case the oscillations eventually ``die out'' after 5950 \yr{} (Fig.~\ref{fig:anisotropic_sloshing}b, green line). Using a slightly larger time-step size results in growth of the amplitude of the oscillations with each time step, until the numerical method eventually diverges. Figure~\ref{fig:anisotropic_sloshing}c (orange line) shows the surface at 1725 \yr{}, just prior to its divergence. Rather than making the surface slosh around the equilibrium for unstable time-step sizes the instabilities in this case show up as spurious oscillations in the surface. 

Despite the different characters of the instabilities on the isotropic and anisotropic domain, using FSSA ($\theta=1$) leads to a stable solution in both cases (Fig.~\ref{fig:isotropic_sloshing}d using $\Delta t = 0.11$ \yr{} and Fig.~\ref{fig:anisotropic_sloshing}d using $\Delta t = 100$ \yr{}, respectively).


\subsection{Experiment 1.2: The effect of mesh resolution and choice of Euler method}
\subsubsection{Setup}

To further investigate capabilities of FSSA, the dependence of the LST on mesh resolutions and choice of Euler method is presented. Pure sloshing instabilities are expected to be independent of mesh size due to the LST estimate Eq.~\eqref{eq:tau_min}. Such instabilities are also expected to be independent of the choice of forward or backward Euler as the choice of Euler method does not impact the velocity coefficient in the discretized free-surface equation~\eqref{eq:discreteicesurface}. 

In this experiment, the LST is numerically determined for different mesh resolution for both forward and backward Euler. For the isotropic domain the mesh varies using resolutions $(n_x,n_z)=$ $ (10,10),$ $(20,20),$ $(30,30),$ $(40,40),$ $(50,50)$, while for the anisotropic domain it varies with resolutions $(n_x,n_z)$=$(30,3),$ $(50,5),$ $(100,10),$ $(150,15),$ $(200,20)$.

\subsubsection{Results \& Discussion}

 The results are summarized in Table~\ref{tab:LST_vs_mesh}. For the isotropic case the sloshing instability is by all practical means mesh independent. Looking at the anisotropic case in Table~\ref{tab:LST_vs_mesh}, the LST on the contrary has a clear mesh dependence, indicating that the character of the instabilities are different from those on the isotropic domain. 

\begin{table}[H]
		\centering
		\caption{Numerically evaluated LST with and without FSSA, for different mesh resolutions $(n_x,n_z)$ on a geometrically isotropic domain (left) and on an anisotropic domain (right). The second column in each table corresponds to forward Euler (FE) and the third column to backward Euler (BE).}
\label{tab:LST_vs_mesh}
\begin{tabular}{lllllll}
		\multicolumn{3}{c}{Isotropic, no FSSA}                                          &  & \multicolumn{3}{c}{Anisotropic, no FSSA}                                  \\
		\multicolumn{1}{l|}{$(n_x, n_z)$} & \multicolumn{1}{l|}{FE, $\theta=0$}    & BE, $\theta=0$   &  & \multicolumn{1}{l|}{$(n_x, n_z)$} & \multicolumn{1}{l|}{FE, $\theta=0$} & BE, $\theta=0$\\ \cline{1-3} \cline{5-7} 
		\multicolumn{1}{l|}{$(10, 10)$}   & \multicolumn{1}{l|}{0.111 \yr{}} & 0.117 \yr{}&  & \multicolumn{1}{l|}{$(30, 3)$}    & \multicolumn{1}{l|}{90 \yr{}} & 90 \yr{} \\
		\multicolumn{1}{l|}{$(20, 20)$}   & \multicolumn{1}{l|}{0.109 \yr{}} & 0.115 \yr{}&  & \multicolumn{1}{l|}{$(50, 5)$}    & \multicolumn{1}{l|}{79 \yr{}} & 79 \yr{} \\
		\multicolumn{1}{l|}{$(30, 30)$}   & \multicolumn{1}{l|}{0.109 \yr{}} & 0.115 \yr{}&  & \multicolumn{1}{l|}{$(100, 10)$}  & \multicolumn{1}{l|}{72 \yr{}} & 73 \yr{} \\
		\multicolumn{1}{l|}{$(40, 40)$}   & \multicolumn{1}{l|}{0.109 \yr{}} & 0.115 \yr{}&  & \multicolumn{1}{l|}{$(150, 15)$}   & \multicolumn{1}{l|}{65 \yr{}} & 66 \yr{} \\
		\multicolumn{1}{l|}{$(50, 50)$}   & \multicolumn{1}{l|}{0.109 \yr{}} & 0.115 \yr{}&  & \multicolumn{1}{l|}{$(200, 20)$}  & \multicolumn{1}{l|}{62 \yr{}} & 62 \yr{}\\  \cline{1-3} \cline{5-7} 
		\multicolumn{1}{l}{}   & \multicolumn{1}{l}{} & &  & \multicolumn{1}{l}{}    & \multicolumn{1}{l}{} & \\ 
		\multicolumn{1}{l}{}   & \multicolumn{1}{l}{} & &  & \multicolumn{1}{l}{}    & \multicolumn{1}{l}{} & \\ 
		\multicolumn{3}{c}{Isotropic, FSSA}                                          &  & \multicolumn{3}{c}{Anisotropic, FSSA}                                  \\
		\multicolumn{1}{l|}{$(n_x, n_z)$} & \multicolumn{1}{l|}{FE, $\theta=1$}    & BE, $\theta=1$   &  & \multicolumn{1}{l|}{$(n_x, n_z)$} & \multicolumn{1}{l|}{FE, $\theta=0$} & BE, $\theta=0$\\ \cline{1-3} \cline{5-7} 
		\multicolumn{1}{l|}{all}   & \multicolumn{1}{l|}{1 \yr{}} & 1 \yr{}&  & \multicolumn{1}{l|}{all}    & \multicolumn{1}{l|}{1000 \yr{}} & 1000 \yr{} \\  \cline{1-3} \cline{5-7}
\end{tabular}
\end{table}

For both the isotropic and anisotropic case, a backward Euler discretization is only slightly more stable than a forward Euler discretization. This demonstrates that the sloshing instability can not be remedied by simply using a semi-implicit scheme (i.e., backward Euler) to discretize the free-surface equation. 

Applying FSSA stabilizes the problem for all mesh resolutions on both the anisotropic and isotropic domain, with both forward and backward Euler. The FSSA method thus also seems to stabilize the mesh-dependent oscillations at the surface, appearing on the anisotropic domain.The largest stable time-step sizes tested using FSSA were as large as $\Delta t = 1$ \yr{}} (isotropic case) and $\Delta t = 1000$ \yr{} (anisotropic case).


\subsection{Experiment 2: Two-dimensional moving-margin experiment}
\subsubsection{Setup}\label{sec:eismintsetup}

\begin{figure}[H]
\centering
\includegraphics[width=\linewidth]{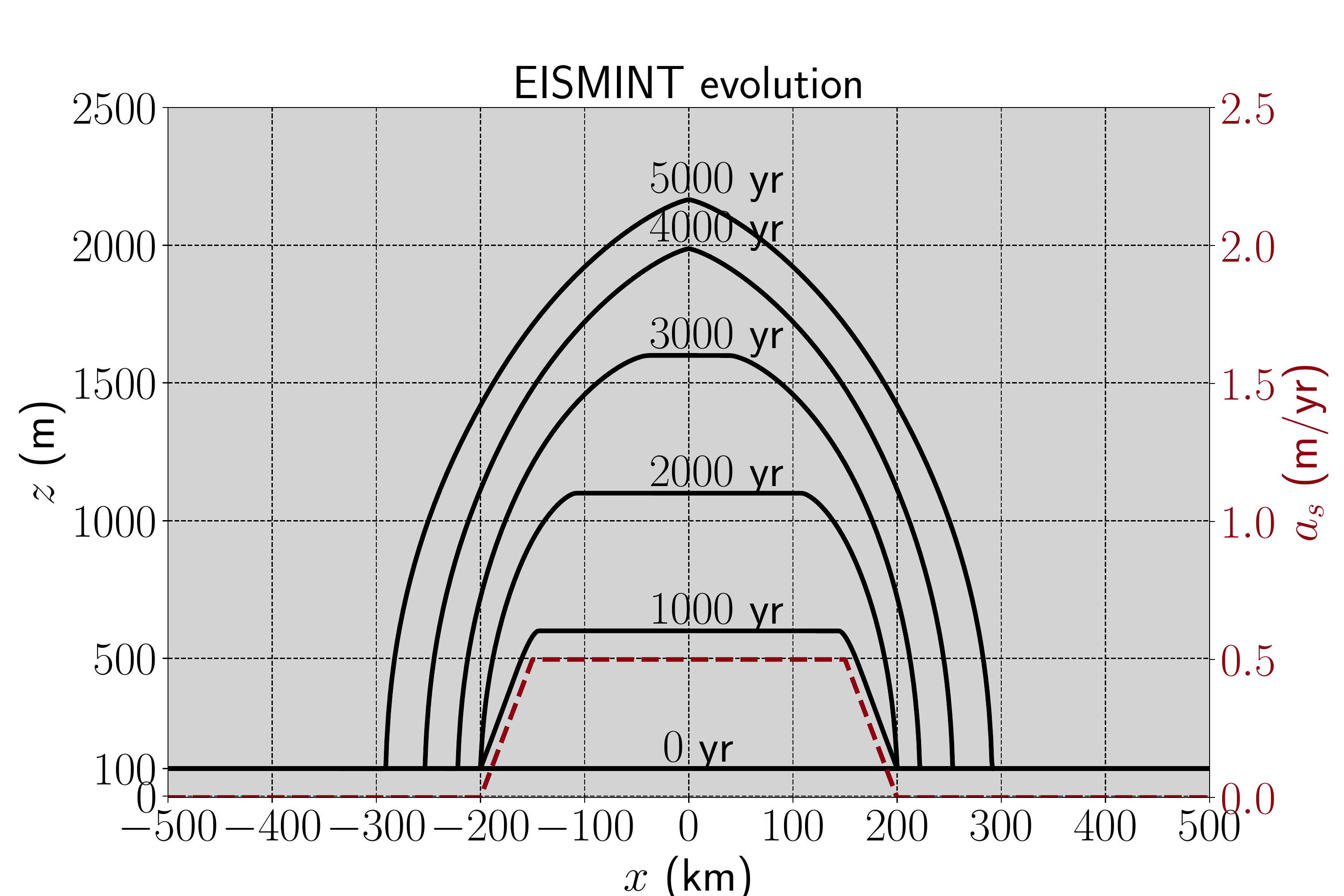}
\caption{Surface profiles of an evolving, initially flat, ice sheet (black solid lines) subject to an accumulation rate given by Eq.~\eqref{eq:accumulation_eismint} (red dashed line). }
\label{fig:EISMINT-evolution}
\end{figure}

To study the instabilities and effect of FSSA on a scenario relevant for long-term prognostic ice-sheet simulations, the build up of an ice sheet with moving margins is simulated, using the same set up as in \citep{CHENG201729}. The experiment is inspired by the EISMINT moving-margin benchmark experiment \citep{Huybrechts}. The simulation starts from a flat domain with a thickness of 100 m and horizontal extent $L_x=1000$ km, subject to the accumulation rate 
\begin{equation}
		a_s(x) = \max(0, \min(s_0, s(R - 0.5 L_x))),
\label{eq:accumulation_eismint}
\end{equation}
where $R=200$ km is the radius of the region of accumulation, $s_0=5\times10^{-4}$ m/\yr{} is the maximum accumulation rate and $s = 10^{-5}$ \yr$^{-1}$ (red dashed line in Fig.~\ref{fig:EISMINT-evolution}). The mass accumulation is concentrated to a region in the interior, yielding a surface gradient resulting in a flow of ice directed away from the center. If melting would occur at the ice-sheet margins, for example due to a warmer climate in lower altitudes
, the system would at some point reach a steady state. 
However, this experiment does not consider melting and consequently 
the ice-sheet margin moves away from the center indefinitely (see the black lines in Fig.~\ref{fig:EISMINT-evolution}).

The goals of this experiment are to: 1) study how the LST evolves as the ice sheet builds up, with and without FSSA, 2) measure the error potentially introduced by FSSA and how it depends on the stabilization parameter $\theta$, 3) compare the LST for the FSSA method applied to forward Euler and backward Euler.  

To automatically estimate the LST over time, the backtracking stability-estimate algorithm in Alg.~\ref{alg:stability-est} is used. In summary, the algorithm slowly increases the time-step size $\Delta t$ until it is so large that the simulation would becomes unstable in subsequent time steps, at which point it is cut in half. To check for stability issues, a test simulation of five extra time steps is done. These five extra steps which are done in each time step are then discarded. 
\begin{algorithm}[H]
\caption{Backtracking stability-estimate algorithm}
\label{alg:stability-est}
\begin{algorithmic}[1]
\State Choose an initial time-step size $\Delta t$
\For {each time step $k$}
\State Do five test steps using $\Delta t$
\State Check if the solution is stable, e.g., $|| \frac{dh}{dx} || < TOL$
\If{stable}
	\State Advance one time step, $t = t + \Delta t$
	\State Increase the time-step size by 5 \%, $\Delta t = 1.05 \Delta t$
\Else
	\State Halve the time-step size, $\Delta t = 0.5 \Delta t$
	\State Go back to (3)
\EndIf
\EndFor
\end{algorithmic}
\end{algorithm}
The stability of the solver is determined by the condition $|| \frac{dh}{dx} ||_{\infty} < TOL $, with $TOL = 0.5 \times 10^{-3}$. The tolerance is chosen to be five times as large as the maximum surface gradient observed over the simulation period (5000 \yr{}) for a stable solution. The mesh resolution for this case is $(n_x,n_z)=(400,10)$.

The accuracy of FSSA is measured by comparing surface profiles to a numerical reference solution. To ensure that spatial errors do not pollute the error measurement, a very fine resolution of $(n_x,n_z)=(1600,10)$ is used for all simulations related to the error calculations. 

The error calculations are done for a surface profile starting at 5000 \yr{}. A reference solution $h^*$ is then constructed by performing a simulation from 5000 \yr{} to 5100 \yr{}, using a small time-step size $\Delta t= 0.1$ \yr{} with forward Euler discretization and no FSSA ($\theta=0)$. The buildup of the ice sheet is done using forward Euler with FSSA. To speed up the initial calculation, the time-step size and the stabilization parameter are gradually decreased from $\Delta t=10$ \yr{} to $\Delta t=1$ \yr{} and $\theta=1$ to $\theta=0$, respectively. The simulation from 5000 \yr{} to 5100 \yr{} is performed using a time-step size $\Delta t=1$ \yr{}, varying $\theta$ for both forward and backward Euler. The error is then computed at the end of the simulation period ($5100$ \yr{}) and is measured in the $L^1$-norm, according to 

\begin{equation}
  \epsilon_h = \frac{1}{|\Omega_h|} \int_{\Omega_h} |h - h^*|\ dx,
  \label{eq:average_error}
\end{equation} 
where $\Omega_h$ is the horizontal region bounded by the ice-sheet margins. This region increases over time due to the movement of the margins. 

\subsubsection{Results \& Discussion}

\begin{figure}[H]
\centering
	\includegraphics[width=\linewidth]{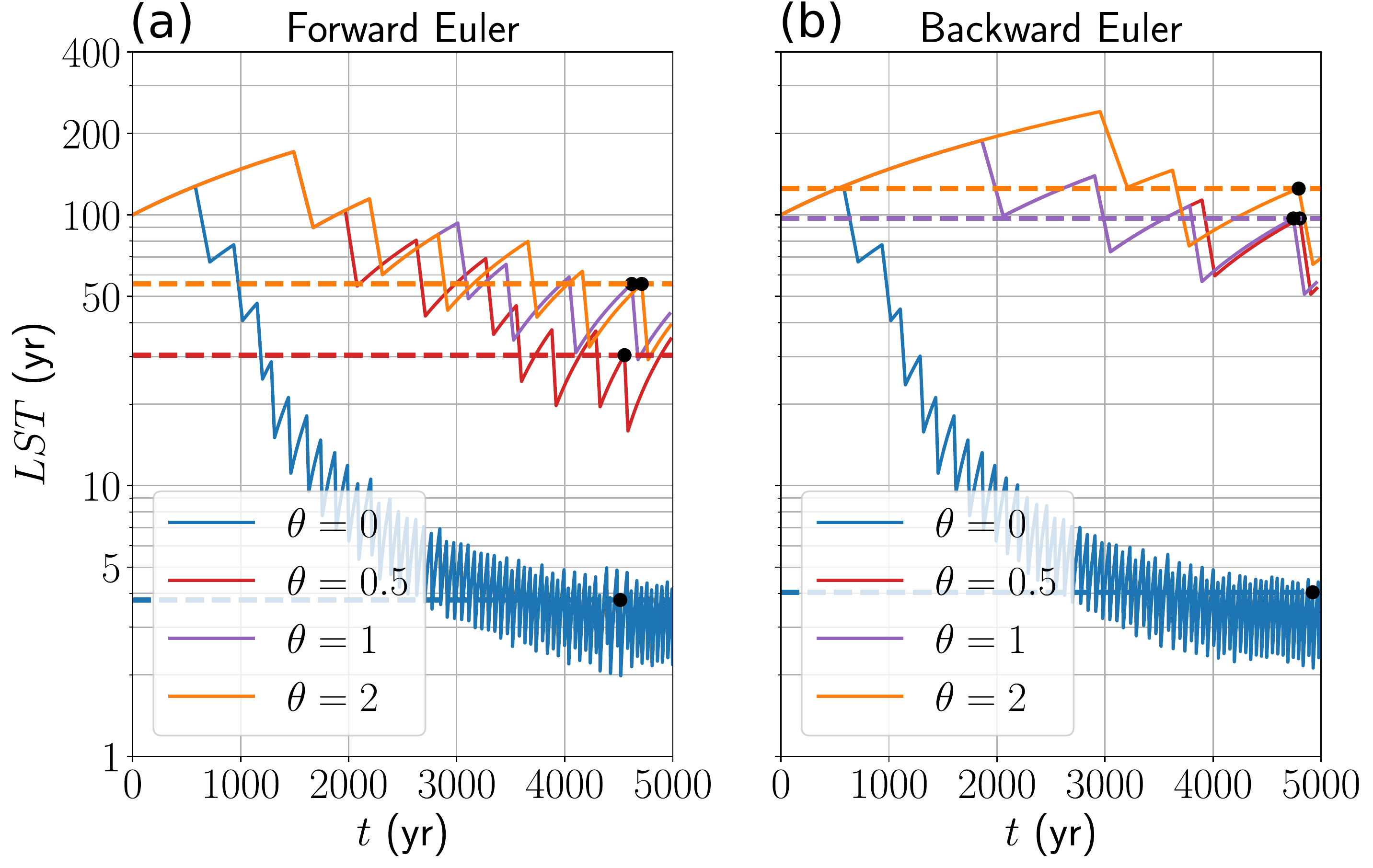}
		\caption{Estimation of the LST on EISMINT over time for the unstabilized problem, i.e., stabilization parameter $\theta=0$, and the problem stabilized using FSSA ($\theta=0.5,1,2$). The free-surface equation~\eqref{eq:discreteicesurface} is discretized using either (a) forward Euler or (b) backward Euler. The black dots and the corresponding dashed lines mark the estimated LST over the whole simulation period.}
\label{fig:EISMINT_stability}
\end{figure}

Fig.~\ref{fig:EISMINT_stability} shows how the LST evolves over time for different values of the stabilization parameter $\theta$ and Euler method. 

For all cases, the LST is seen to decrease over time. This is expected because all potentially relevant time-step size restrictions (i.e., the time-step size restriction of Eq.~\eqref{eq:tau_min}, as well as a classical CFL condition, or potentially the time-step size restriction of the shallow-ice approximation mentioned in the introduction) are inversely proportional to the velocity. As the ice sheet evolves its margins steepen resulting in an increase of the velocity field, thereby reducing the stable time-step size. 

The sawtooth-like pattern exhibited in the general decrease of the LST (Fig.~\ref{fig:EISMINT_stability}) is an expected consequence of the backtracking algorithm.

To quantify the LST in this case, it is defined to be the largest known stable time-step size over the whole simulation period. These are the values of the time-step sizes at the peaks marked with black dots in Fig.~\ref{fig:EISMINT_stability}. It is also seen in Fig.~\ref{fig:EISMINT_stability} 
that the numerical method is substantially more stable with FSSA when compared to using backward Euler or forward Euler alone. 

\begin{table}[H]
		\caption{Measured LST without and with FSSA using stabilization parameters $\theta=0.5,1,2$. The free-surface equation~\eqref{eq:discreteicesurface} is discretized using either forward Euler (FE) or backward Euler (BE). For the problem stabilized with FSSA, the relative increase in the time-step size compared to the unstabilized problem, $LST(\theta\neq0)/LST(\theta=0)$, is reported.}
\begin{center}
\begin{tabular}{lccc}
Stabilization & Euler Method & LST (\yr{}) & $LST(\theta)/LST(0)$  \\
    \hline
        \hline
    \multirow{2}{*}{Unstabilized ($\theta=0$)}         &FE& 3.8 & - \\
    										           &BE& 4.0 & - \\
        \hline
    \multirow{2}{*}{FSSA, $\theta=0.5$}    &FE& 30 & 8.0 \\
    										           &BE& 97 & 24.0 \\
        \hline
    \multirow{2}{*}{FSSA, $\theta=1$}      &FE& 56 & 14.7 \\
    												   &BE& 97 & 24.0 \\
    \hline
    \multirow{2}{*}{FSSA (over-stabilized), $\theta=2$}&FE& 56 & 14.7 \\
    												   &BE& 125 & 30.9 \\
    \hline
\end{tabular}
\end{center}
\label{tab:EISMINT_tab}
\end{table}

The obtained LST for different values of $\theta$ is presented in Table~\ref{tab:EISMINT_tab}, where the relative increase in the LST compared to no stabilization is also presented. Table~\ref{tab:EISMINT_tab} shows that the least stable solver is forward Euler without FSSA, followed by backward Euler without FSSA. The most stable solver is backward Euler with $\theta=2$, i.e., over-stabilized. 

Contrary to what was observed for the sinusoidal domain in Experiment 1, using FSSA combined with backward Euler is far more stable than using FSSA combined with forward Euler, in the experiment considered. Interestingly, comparing forward and backward Euler without the use of FSSA, reveals that backward Euler is now only slightly more stable than forward Euler. 

\begin{table}[H]
		\caption{Measured error $\epsilon_h$ without and with FSSA using $\theta=0.5,1,2$. The free-surface equation~\eqref{eq:discreteicesurface} is discretized using either forward Euler (FE) or backward Euler (BE). The error $\epsilon_h$ is measured at the end of simulations which ran for 100 \yr{} (starting at 5000 \yr{} and ending at 5100 \yr{}). The error is evaluated according to Eq.~\eqref{eq:average_error} using the reference solution $h^*$ described in Sect.~\ref{sec:eismintsetup}. All simulations used a time-step size $\Delta t=1$.}
\begin{center}
\begin{tabular}{lcc}
Stabilization & Euler Method & $\epsilon_h$ (m)  \\
    \hline
        \hline
    \multirow{2}{*}{Unstabilized ($\theta=0$)}         &FE& $\num{0.157}$ \\
    										           &BE& $\num{0.137}$ \\
        \hline
    \multirow{2}{*}{FSSA stabilized ($\theta=0.5$)}    &FE& $\num{0.171}$  \\
    										           &BE& $\num{0.151}$ \\
        \hline
    \multirow{2}{*}{FSSA stabilized ($\theta=1$)}      &FE& $\num{0.178}$ \\
    												   &BE& $\num{0.159}$ \\
    \hline
    \multirow{2}{*}{FSSA (over)stabilized ($\theta=2$)}&FE&$\num{0.186}$  \\
    												   &BE& $\num{0.166}$ \\
    \hline
\end{tabular}
\end{center}
\label{tab:EISMINT_error}
\end{table}

Table~\ref{tab:EISMINT_error} shows how $\theta$ for forward Euler and backward Euler impacts the error (calculated as described in Sect.~\ref{sec:eismintsetup}). The error introduced by FSSA is only marginally larger than the error of the unstabilized problem. A slight increase of the error is seen for an increasing $\theta$. 

The stabilization does not measurably increase memory usage or the computation time per nonlinear iteration. However, to reach the same nonlinear tolerance, the FSSA method did require about 10-20 \% more Picard iterations for $\theta \ge 0.5$. Since computation times are proportional to the number of nonlinear iterations, the speedup gained is therefore approximately $1/1.2 \times LST(\theta)/LST(0)$. For the most stable cases this corresponds to a speedup of $12.3$ and $52.2$ for forward and backward Euler respectively.


\subsection{Experiment 3: Evolution of a three-dimensional Vialov profile}
\subsubsection{Setup}
\label{sec:vialovsetup}

This experiment investigates the performance of FSSA when applied to a 3D ice sheet. The set-up is similar to the previous EISMINT experiment (Experiment 2); however, to reduce the computational effort, rather than starting from an incipient ice sheet (as in Experiment 2), the analytical the Vialov profile \citep{vialov1958}, common in glaciology, is used as a starting point. It is given by

\begin{equation}
  h(r) = z_0 + h_0 \left [ 1 - \left ( \frac{r}{r_0} \right )^{\frac{n+1}{n}}  \right ]^{\frac{n}{2n+2}},
\end{equation}
where $r = \sqrt{x^2 + y^2}$, $h_0 = 3.575$ km is the maximum initial height, $r_0 = 750$ km the radius of the profile, $n=3$ is the stress exponent from Glen's flow law, Eq.~\eqref{eq:glen}, and $z_0 = 0.1$ km is minimum thickness used to prevent the ice sheet from having zero thickness at the front.

To introduce a significant perturbation to the ice sheet the accumulation rate specified in \citet{Ahlkrona2016} is used. The accumulation rate is

\begin{equation}
	a(r) = \min(0.5e-3, 1e-5 (R - r))),
\end{equation}
where $R = 450$ km is the radius of the region with accumulation: interior regions of the ice-sheet domain exhibit a positive mass balance while regions closer to the front exhibit an increasingly negative mass balance. Since the initial profile is far from steady state, the model is spun up by a $10$-year long transient simulation, using forward Euler with a time-step size $\Delta t = 0.1$ \yr{}. From the resulting slightly relaxed surface, a suite of simulations using forward Euler without FSSA ($\theta = 0$) and with FSSA ($\theta = 1$) are run to evaluate the size of the stable time steps for the two different cases. The forward Euler method is chosen primarily due to its simplicity and low computational cost. In addition, forward Euler is generally more unstable than other time-stepping schemes, thus the argument is made that it can be used to establish as a lower bound on the potential increase in stability that may result from using FSSA. 

All simulations are run for $20$ \yr{}, at the end of which the resulting surfaces are compared to a reference solution using the error measure in Eq.~\eqref{eq:average_error}. The reference solution, $h^*$, is calculated using a short time-step size $\Delta t = 0.05$ \yr{} for which the transient solution exhibits stability. Due to the steep fronts of the ice-sheet margin, this is typically where oscillations initiate and result in negative surface heights. Therefore, in this experiment a stable solution is defined to be an evolution of the surface such that negative surface heights are never encountered throughout the simulation. Due to the relatively steep ice-sheet margins, a horizontal footprint mesh that is refined in the radial direction towards the margins is used. The radial resolution is finest at the ice-sheet margin (3.45 km) and largest in the interior of the ice sheet (approximately 60 km); see Fig.~\ref{fig:vialov_diff}. The final 3D mesh is the result of vertically extruding the horizontal footprint in 10 layers, giving 115740 elements. 

\subsubsection{Results \& Discussion}
\label{sec:vialovresults}
Table~\ref{tab:3d_surface_comparison} shows how the error changes with increasing time-step sizes, $\Delta t$, using forward Euler time discretization without (unstabilized) and with FSSA. The LST for forward Euler without FSSA is $\Delta t = 0.2$ \yr{}. When the FSSA is used, it is seen that forward Euler is stable for at least time-step sizes up to $\Delta t = 10$ \yr{} (the largest time-step size tested) under the whole simulation period of 20 \yr{}. The error is proportional to $\Delta t$, as expected from a forward Euler discretization. This demonstrates that any error introduced by FSSA is small in comparison to the error that stems from the underlying time discretization of the free-surface equation.

\begin{figure}[H]
\centering
	\includegraphics[width=0.8\linewidth]{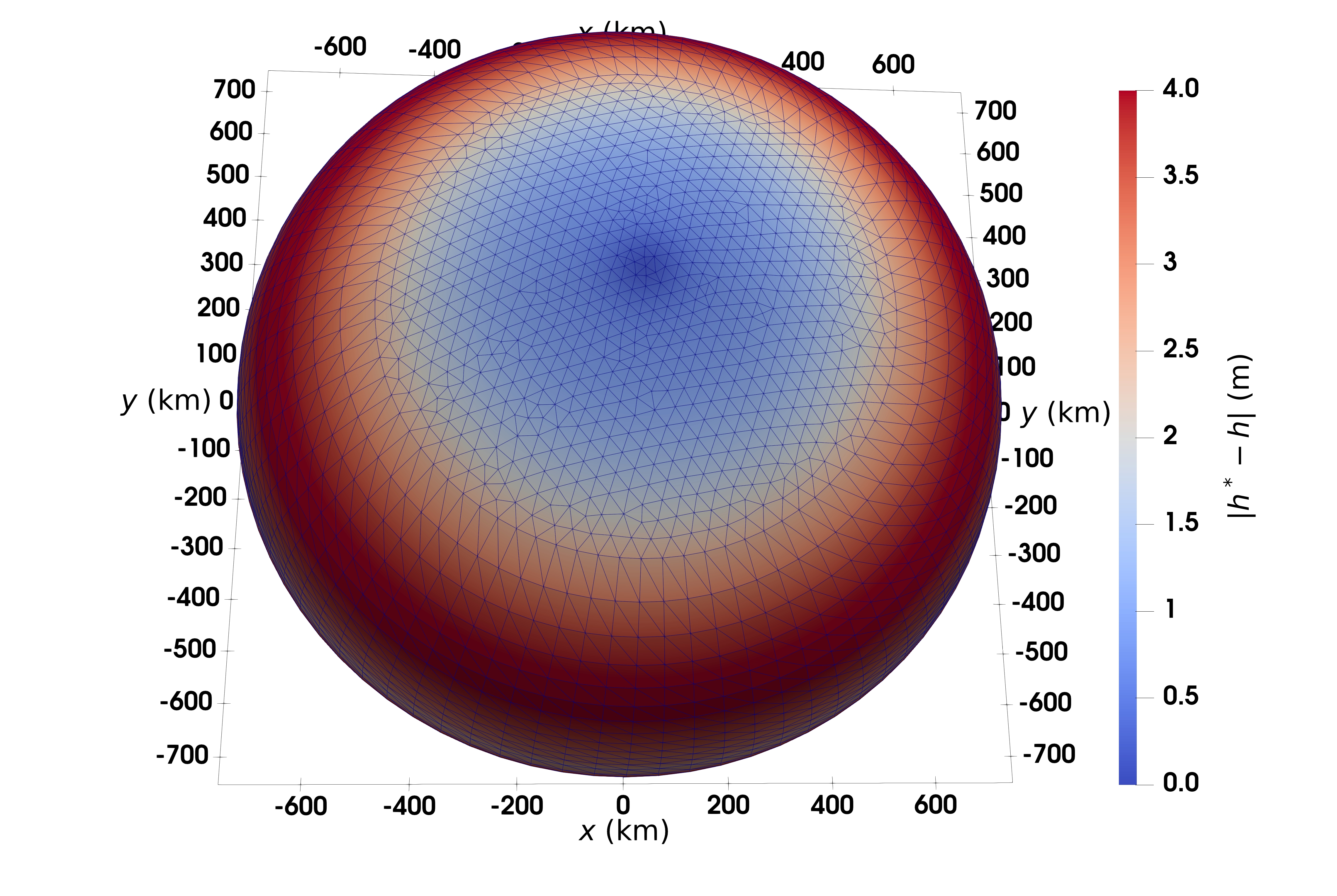}
		\caption{Difference in surface height of the Vialov-profile between the reference solution, $h^*$, and a simulation, $h$, using FSSA. The surface of the ice sheet is evaluated after $30$ \yr{} of total simulation time (running for $20$ \yr{} from the relaxed surface). The color shows $|h^* - h|$ in meters at the end of the simulation, i.e., the absolute difference between the reference solution (calculated using forward Euler with $\Delta t = 0.05$ \yr{}) and the surface resulting from using forward Euler and FSSA, with a time-step size $\Delta t = 10$ \yr{}.}
\label{fig:vialov_diff}
\end{figure}

Figure~\ref{fig:vialov_diff} shows the absolute difference between the reference solution and the solution stabilized with FSSA using $\Delta t = 10$ \yr{} at the final simulation time $t = 20$ \yr{}.

\begin{table}[H]
\caption{
  Measured error $\epsilon_h$ without and with FSSA for Experiment 3. The error $\epsilon_h$ for different time-step sizes $\Delta t$ is measured at the end of simulations which run for 20 \yr{} from a relaxed Vialov surface. The error is evaluated according to Eq.~\eqref{eq:average_error} using the reference solution $h^*$ described in Sect.~\ref{sec:vialovsetup}. All simulations use the forward Euler method for the discretized free-surface equation~\eqref{eq:discreteicesurface}.
  }
  \label{tab:3d_surface_comparison}
  \centering
  \begin{tabular}{lcc} 
    Stabilization & $\Delta t$ (\yr{}) & $\epsilon_h$ (m)\\
    \hline \hline
    \multirow{2}{*}{Unstabilized ($\theta=0$)}
                  & \num{0.1} & $\num{0.030}$ \\
                  & \num{0.2} & $\num{0.092}$ \\
    \hline
    \multirow{7}{*}{FSSA, $\theta=1$}
                  & \num{0.1}  & \num{0.059} \\
                  & \num{0.2}  & \num{0.087} \\
                  & \num{0.5}  & \num{0.172} \\
                  & \num{1}  & \num{0.314} \\
                  & \num{2}  & \num{0.597} \\
                  & \num{5}  & \num{1.433} \\
                  & \num{10}  & \num{2.793} \\
    \hline
  \end{tabular}

\end{table}

\section{Summary and Conclusions}\label{sec:summary}

In this work a free-surface stabilization algorithm (FSSA) for Stokes-coupled free-surface flow have been adapted to the regime of ice-sheet modeling. It shown numerically that the method increases the largest stable time step by at least an order of magnitude. Previously, FSSA had only been used to stabilize free-surface flow in the regime of mantle convection for linear Stokes-flow models. In addition to ice flow having a nonlinear rheology, ice-sheet domains are also characterized by being highly anisotropic, in the sense that the horizontal extent can be up to 1000 times greater than its vertical extent. Despite these differences, the combined results of the simulations showed that FSSA increased largest stable time-step sizes (LST) substantially, while any approximation error introduced were found to be small relative the error due to the temporal discretization.

Three different experiments were designed to assess the stabilizing impact of FSSA, as well as monitoring any error potentially introduced. In Experiment 1, Sect.~\ref{sec:experiments}, it was verified that isotropic domains indeed give rise to a mesh independent sloshing instability - the type of instability FSSA was originally developed to mitigate. However, when the domain was extruded 100 times in the horizontal direction, the sloshing instability could no longer be observed; instead mesh-resolution-dependent spurious oscillations appeared in the solution. Another difference is that in the isotropic case, a crude theoretical estimate (derived for the sloshing instability) of the largest stable time-step size could be used to predict the LST fairly accurately. However, for the anisotropic domain the same theoretical LST estimate was off by about two orders of magnitude. This discrepancy indicates that the sloshing time-step-size constraint is unsuitable for predicting LST in ice-sheet modeling; developing a theoretical time-step-size constraint for the anisotropic case could potentially be of interest for numerical ice-sheet modeling. 

Despite the differences in behavior of the instabilities for the two domains, FSSA was still able to stabilize the ice-sheet solver, and increased the LST substantially. In this experiment, the choice of Euler discretization for the free-surface equation had a very little to no impact on the stability of the solver. 

Experiment 2, Sect.~\ref{sec:experiments}, looked at how FSSA fared when applied in a more realistic, two-dimensional (2D), scenario, relevant to ice-sheet modeling. It was found also for this case that FSSA mitigated instabilities and increased the LST substantially - at least one order of magnitude. The choice of Euler method for updating the free surface alone also in this experiment had little impact on the LST; however, when FSSA was combined with a backward Euler time-stepping scheme, the solver was the most stable, and further increased the LST by about a factor of two to three. 

This experiment also showed that FSSA stabilized the problem at the slight expense of making the flow more nonlinear, manifested as a minor increase in the nonlinear iterations needed to reach the same tolerance, as without FSSA. 

Another important factor found to affect the stability of FSSA was the stabilization parameter $\theta$. Generally, increasing $\theta$ led to a more stable solver; however, it was demonstrated that increasing $\theta$ comes at a slight cost of increasing the error in the solution. Thus, there is a trade-off between stability and accuracy. For this case, a $\theta$ value between 0.5 and 1.0 was a good choice. If stability is of highest priority a larger value of $\theta$, i.e. over stabilization, is a better choice.

Experiment 3, Sect.~\ref{sec:experiments}, considered FSSA applied to a three-dimensional (3D) case, also in an ice-dynamical setting. It was shown for this case as well, that FSSA made the solver more stable; similarly to Experiment 2, FSSA increased stable time-step sizes by at least one order of magnitude. However, due to the computational cost of performing simulations in three-dimensions, these simulations were only done using forward Euler discretization; it remains to assess the stabilizing impact of FSSA for backward Euler in 3D, or higher order time-stepping schemes for both 2D and 3D.

These experiments demonstrated that FSSA has a great potential in increasing stable time-step sizes in ice-sheet simulations. The goal is ultimately to increase stable time-step sizes for real-world ice sheets, i.e., Antarctica and Greenland. However, such problems are much more complex than the experimental set-ups presented here. A first step toward realizing this could be to analyze FSSA when more physical processes are included in the ice-sheet model, e.g., temperature-dependent viscosity, basal sliding or taking ice/ocean interactions into account.

\section*{Acknowledgements}
The authors acknowledge the funding provided by the Swedish e-Science Research Centre (SeRC).

\bibliographystyle{abbrvnat}

\end{document}